**Unveiling hidden ferrimagnetism and giant magnetoelectricity in polar magnet $Fe_2Mo_3O_8$**

Yazhong Wang[1], Gheorghe L. Pascut[1], Bin Gao[1], Trevor A. Tyson[1,2], Kristjan Haule[1], Valery Kiryukhin[1], and Sang-Wook Cheong[1,*]

[1]Rutgers Center for Emergent Materials and Department of Physics and Astronomy, Rutgers University, Piscataway, New Jersey 08854, USA

[2]Department of Physics, New Jersey Institute of Technology, Newark, New Jersey 07102, USA

[*]Corresponding author, sangc@physics.rutgers.edu

Magnetoelectric (ME) effect is recognized for its utility for low-power electronic devices. Largest ME coefficients are often associated with phase transitions in which ferroelectricity is induced by magnetic order. Unfortunately, in these systems, large ME response is revealed only upon elaborate poling procedures. These procedures may become unnecessary in single-polar-domain crystals of polar magnets. Here we report giant ME effects in a polar magnet $Fe_2Mo_3O_8$ at temperatures as high as 60 K. Polarization jumps of 0.3 $\mu C/cm^2$, and repeated mutual control of ferroelectric and magnetic moments with differential ME coefficients on the order of $10^4$ ps/m are achieved. Importantly, no electric or magnetic poling is needed, as necessary for applications. The sign of the ME coefficients can be switched by changing the applied "bias" magnetic field. The observed effects are associated with a hidden ferrimagnetic order unveiled by application of a magnetic field.

**Introduction**

A significant effort has been invested into finding new materials in which macroscopic properties, such as the magnetization and the electric polarization, are coupled and controlled by external parameters like temperature and electric or magnetic fields[1-5]. Materials where these quantities are interconnected are highly desired due to their importance in developing devices with new functionalities[6-8]. Examples of materials falling into this category are the pyroelectric and multiferroic materials[9-12]. The practical aspect of pyroelectric materials is the capacity to generate a current when they are subjected to a temporal temperature gradient through heating or cooling. Due to the efficient conversion of thermal energy into electrical energy, pyroelectric materials have offered numerous device applications, for example for temperature-sensing[13,14] and for thermoelectric applications[15]. The practical aspect of multiferroic materials is the ability to mutual control the magnetization (polarization) by the use of external electric (magnetic) fields, the effect known as magnetoelectric (ME) effect[16-18]. The ME effect can be linear or/and non-linear with respect to the external fields and it is characterized by the appropriate ME coefficients[19,20]. At the present time, materials with large ME coefficients are exploited for developing low-power magnetoelectronic-based devices and new multiple state memory elements[21,22]. Recently, materials with significant ME response associated with ferroelectricity induced by magnetic order have been identified[23-25]. Unfortunately, elaborate poling procedures, such as cooling in applied electric and magnetic fields, are needed to reveal the largest ME coefficients in these systems. Finding new materials with colossal ME coefficients lacking this drawback is of primary importance for prospective applications.

Materials belonging to the polar crystallographic symmetry groups lack the inversion symmetry at all temperatures. Many of these materials contain magnetic ions, and they often exhibit long-range magnetic order. We call these materials "polar magnets". The prerequisite for non-trivial magnetoelectricity is simultaneous breaking of time reversal symmetry and space inversion symmetry. Thus, all polar magnets should exhibit non-trivial ME effects below magnetic ordering temperatures. Importantly, monodomain polar single crystals can often be grown, potentially eliminating the need for any poling procedures to reveal the largest possible ME response. While the polar magnets are numerous, the investigation of their magnetoelectricity has been extremely limited. The few examples of polar magnets whose magnetoelectricity has been studied include $GaFeO_3$ (ref. 26) and $Ni_3TeO_6$ (ref. 24). Clearly, a targeted search for enhanced ME effects among polar magnets holds significant promise.

For the ME device applications, ferro- or ferrimagnetic polar magnets are some of the best candidates, as macroscopic magnetic moment is needed for their functionality. Such compounds are rare. However, in some cases macroscopic magnetic moment is "hidden" within a nominally antiferromagnetic state, and can be easily revealed in a modest applied magnetic field, thereby leading to a potentially large ME response. A well-known example of such a hidden moment is realized in $La_2CuO_4$, the parent compound of high-$T_C$ cuprate superconductors[27]. Each Cu-O plane exhibits a weak ferromagnetic moment due to canting of the spins of the otherwise regular Neel order. The canting results from Dzyaloshinsky-Moria interaction. Weak ferromagnetism is masked in zero magnetic field because of the antiferromagnetic interplane coupling. However, spin canting is responsible for the many distinct magnetic properties of this compound, including the unusual shape of the magnetic susceptibility in the vicinity of $T_N$ and in an applied field, and the atomic-scale giant magnetoresistence in the field-induced weakly ferromagnetic phase. Another layered magnet in which a small ferrimagnetic moment of each layer is hidden at zero field is multiferroic (but nonpolar at high $T$) $LuFe_2O_4$ (ref. 28). The giant magnetic coercivity and the unusual ME relaxation properties of $LuFe_2O_4$ are related to the ferrimagnetism in its Fe-O layers. Similar to these compounds, a hidden magnetic moment in a polar magnet could result in a strongly enhanced magnetic response, which should lead to large ME effects when the magnetic moment and crystal structure are coupled.

Herein, we report giant ME effects in a monodomain polar magnet $Fe_2Mo_3O_8$, that possesses both multiferroic and pyroelectric characteristics. Below $T_N \approx 60$ K, it exhibits a layered collinear magnetic structure with a small ferrimagnetic moment in each layer[29]. As in $La_2CuO_4$, this moment is "hidden", but can be revealed in a modest applied magnetic field[28]. As a result of field- and temperature-induced magnetic transitions in $Fe_2Mo_3O_8$, the electric polarization exhibits changes as large as 0.3 $\mu C/cm^2$. As the hidden ferrimagnetism is converted to a bulk moment by an applied magnetic field, giant differential ME coefficients approaching $10^4$ ps/m are achieved. The observed effects are significantly larger than those previously reported in polar magnets, such as $Ni_3TeO_6$ (ref. 24). The ME control is mutual, as both the magnetization and electric polarization can be tuned by the electric and magnetic field, respectively. Importantly, no electric or magnetic poling is needed, and the sign of the differential ME coefficients can be switched by simply changing the applied "bias" magnetic field. Using first principles calculations, we show that exchange striction is the leading mechanism responsible for the observed ME effects. Our results demonstrate the promise of polar magnets as ME systems, and indicate that their functional properties could be further enhanced by presence of a local ("hidden") magnetic moment that can be easily converted to macroscopic magnetization by an applied field.

**Results**

$Fe_2Mo_3O_8$, known as the mineral kamiokite[30,31], consists of honeycomb-like Fe-O layers separated by sheets of $Mo^{4+}$ ions, See Fig 1(a). The layers are stacked along the *c* axis. The Fe-O layer is formed in the *ab* plane by corner-sharing $FeO_4$ tetrahedra and $FeO_6$ octahedra, as shown in Fig 1(b). In this layer, the tetrahedral ($Fe_t$) and octahedral ($Fe_O$) triangular sublattices are shifted along the *c* axis by 0.614 Å with respect to each other[31], leading to short and long interlayer Fe-Fe distances, see Fig 1(a). The vertices of the $FeO_4$ tetrahedra point along the positive *c* axis, reflecting the polar structure of $Fe_2Mo_3O_8$ (ref. 31). The Mo kagome-like layer is trimerized. The Mo trimers are in the singlet state, and do not contribute to magnetism[32]. Below $T_N \approx 60$ K, the $Fe^{2+}$ moments exhibit the antiferromagnetic (AFM) order in the honeycomb layers, see Fig 1(c). As discussed below, $Fe_O$ has larger spin than $Fe_t$, and therefore each of the Fe-O layers is ferrimagnetic[33]. Along the *c* axis, the nearest Fe spins are aligned in the same direction, implying ferromagnetic interlayer coupling. The resulting stacking of the ferrimagnetic Fe-O layers along the *c* axis leads to vanishing macroscopic magnetic moment, and we call this state AFM.

The temperature variation of DC magnetic susceptibility $\chi$ in zero field-cooled (ZFC) and field-cooled (FC) processes is shown in Fig 1(e) for the magnetic field both parallel and normal to the *c* axis. The shapes of the curves are consistent with the transition to the AFM order shown in Fig 1(c) at $T_N=61$ K, with $Fe^{2+}$ spins pointing along the *c* axis. The large difference between the *c*-axis and in-pane susceptibilities in the paramagnetic state demonstrates appreciable anisotropy of the $Fe^{2+}$ spins. No thermal hysteresis is observed, see Supplementary Fig 1. A large specific heat ($C_P$) anomaly is present at the magnetic transition, see Fig 1(f). To account for the phonon part, the specific heat was fit to a double Debye model for $T>T_N$ (90 to 200 K). The best fit, shown in Fig 1(f), was obtained for the Debye temperatures $\theta_{D1}$=174 K and $\theta_{D2}$=834 K. It fails for $T<T_N$ as it implies a negative magnetic contribution for $T<50$ K. This indicates an additional lattice contribution for these temperatures, suggesting a structural transition associated with the magnetic order. This suggestion is corroborated by the temperature dependence of the dielectric constant $\varepsilon(T)$ and the variation of the electric polarization $\Delta P(T) \equiv P(T)-P(T=120$ K$)$, both along the $c$ axis, shown in Figs 2(a,b). In particular, the jump of $\Delta P$ at $T_N$ clearly indicates simultaneous magnetic and structural transitions. The magnitude of this jump, ~0.3 µC/cm², is larger than the values typically observed in multiferroics, and is the largest measured value in polar magnets, to our knowledge. Importantly, no poling is needed in an already polar material to observe the changes shown in Figs 2(a,b). In particular, $\Delta P$ was measured by integrating the pyroelectric current on warming after cooling down to $T=5$ K in zero electric field (see Supplementary Fig 2 for details). In our measurements, the direction of the $\Delta P$ vector (along or opposite to the positive direction of the $c$ axis defined above) is undetermined. First principles calculations described below indicate that $\Delta P$ points in the positive $c$ direction, hence we adopt this convention here.

Magnetic field ($H$) induces a metamagnetic transition signaled by sharp magnetization ($M$) jumps, see Fig 2(c). It is accompanied by a structural transition indicated by the corresponding jumps in the electric polarization, as shown in Fig 2(d). A small hysteresis is observed in the latter transition. The $\Delta P \equiv P(H)-P(H=0$ T$)$ vector is in the negative $c$ axis direction, and its value at $T=50$ K is roughly twice as small as the $\Delta P$ induced at $T_N$ for $H=0$ T. No poling of any kind is needed. Replacement of Fe with Mn, as well as Zn doping on the Fe site are known to convert the AFM state observed in $Fe_2Mo_3O_8$ into a ferrimagnetic (FRM) state[32,33], in which the AFM order in the Fe-O layers is preserved, but the spins in every second layer are flipped, see Fig. 1(c). In the FRM state, the ferrimagnetic moments of the Fe-O layers are co-aligned, giving rise to a macroscopic magnetization. The extrapolation of the high-field $M(H)$ data of Fig 2(c) to zero field gives a positive intercept of ~0.5 $\mu_B$/f.u. at $T=50$ K, indicating the ferrimagnetic character of the high-field state, which we assume to have the same FRM structure as shown in Fig 1(c). This assumption is corroborated by the Fe-O net ferrimagnetic moment of 0.6 $\mu_B$/f.u. for a single layer, expected from the Moessbauer measurements of the $Fe_O$ and $Fe_t$ moments[33] (4.83 $\mu_B$ and 4.21 $\mu_B$, respectively), as well as by the results of the first principles calculations described below.

**Discussion**

To understand the microscopic origin of the observed ME effects, we have carried out ab-initio calculations in the framework of density functional theory adding an on-site Coulomb self-interacting potential $U$ (DFT+U). For the DFT part, the generalized gradient approximation Perdew-Burke-Ernzerhof (GGA-PBE) functional was used. For $U$=0, the ground state is metallic with the FRM structure, but moderate correlation strength ($U$=4 eV) leads to an AFM insulating ground state. While $U$ of the order of 4 eV is required to obtain the correct ground state, its exact value was found to be unimportant for the magnetic exchange energies relevant to this work. The details of the DFT calculations can be found in the Methods section. The ionic positions were optimized for two imposed magnetic structures, the AFM and FRM. The FRM structure was found marginally higher in total energy (less than 10 meV/f.u.), indicating that this phase is expected to be induced in modest magnetic fields, consistent with our experimental data.

The calculated ionic shifts for the transitions from the paramagnetic (PARA) to the AFM state, and from AFM to FRM, are shown in Figs 1(c), and 3(c,d). The ionic shifts for every ion in the unit cell are given in Supplementary Table I. The experimental paramagnetic structure, and the calculated AFM and FRM structures were used. The ionic shifts can be utilized for an estimate of the magnetically-induced electric polarization change $\Delta P$. While the total polarization is a multivalued quantity, the difference $\Delta P$ between two structures is a well-defined quantity[34]. For a qualitative comparison with experiment, it is sufficient to use the ionic-like formula for $\Delta P$ given by $\frac{1}{V}\sum_j(z_j^f - z_j^i)Q_j$, where $z_j^i$ and $z_j^f$ are the c-axis ionic coordinates for the initial and the final structures, respectively, $Q_j$ are the formal ionic charges, $V$ is the unit cell volume, and the sum is taken over the unit cell. For the PARA to AFM, and AFM to FRM transitions, we obtain $\Delta P$ values of 0.60(11) μC/cm$^2$ and -0.55(11) μC/cm$^2$, respectively. The calculated magnitudes and the relative signs of $\Delta P$ are in good qualitative agreement with our experiments. The positive sign of $\Delta P$ for the PARA to AFM transition indicates that the $\Delta P$ vector points along the positive $c$ axis, justifying the convention used in our work.

The calculated ionic shifts also allow to get an insight into the mechanism of the ME effect. The atoms shift to maximize the magnetic energy gains in the AFM and FRM states. Oxygen ions exhibit the largest shifts, and therefore the ME energy gains should be associated with the modifications of the superexchange paths between the interacting $Fe^{2+}$ spins. Lattice structure, as well as preservation of the in-plane magnetic order in applied magnetic field imply that the largest magnetic coupling ($J$) is between the nearest $Fe^{2+}$ ions, see Fig 1(b). The calculations show that upon the transition from the paramagnetic to the AFM state, the Fe-O-Fe angle ($\theta$) between the nearest $Fe^{2+}$ increases from 109º to ~111º, mostly due to the oxygen shifts, see Fig 1(d). The in-plane antiferromagnetic $J$ increases with increasing $\theta$ due to the more favorable Fe-O-Fe orbital overlap, resulting in the magnetic energy gain. Thus, we ascribe the ionic shifts, as well as the accompanying $\Delta P$, to the exchange striction in the AFM state.

The FRM state can be induced both by a positive and a negative magnetic field along the $c$ axis. The two states differ only by 180º rotation of every spin in the system. While the field-induced magnetizations should be opposite for the opposite fields, $\Delta P$ induced by exchange striction should be identical. This prediction is clearly confirmed by the data of Figs 3(a,b). The calculated ionic shifts for the AFM to FRM transition, shown in Fig 3(d), are opposite (but smaller) to those occurring at the PARA to AFM transition, see Fig 3(c). In other words, the lattice partially relaxes towards the paramagnetic structure in the FRM state. This is consistent with the magnetic energy loss due to the interlayer interactions in the FRM phase, and corresponding relaxation of the lattice distortion realized in the AFM state. As a result, $\Delta P$ is negative in the AFM to FRM transition. Thus, the data of Fig 3(b), in combination with our first principles calculations, show that exchange striction underlies the ME effect in the transition to the FRM state, as in the AFM transition discussed above.

The sharpness of the field-induced transitions shown in Figs 2(d) and 3(a,b) gives rise to giant values of the differential ME coefficient d$P$/d$H$ in the vicinity of the transition field, reaching almost $-10^4$ ps/m for $T$=55 K. (Consult Supplementary Fig 2(d) for the field-dependent d$P$/d$H$ for different temperatures). Combined with absence of poling requirements and the small hysteresis (0.02 T at 55 K, 0.007 T at 58 K), it leads to giant, reproducible, and almost linear variation of $P$ with $H$, as shown in Fig 4(a) for $T$=55 K. In the range shown, $\Delta P$ oscillates, varying by 0.08 $\mu C/cm^2$ as $H$ goes from 3.25 to 3.5 T and back. The inverse effect, in which an applied electric field ($E$) changes the magnetization is also giant, reproducible, and linear, as shown in Fig 4(b). At $T$=55 K and $H$=3.345 T, the magnetization varies by 0.35 $\mu_B$/f.u. in the field oscillating between ±16.6 kV/cm, resulting in the d$M$/d$E$ of -5700 ps/m. Similarly large differential ME coefficients d$P$/d$H$ and d$M$/d$E$ are observed at other points on the AFM-FRM transition boundary shown in Fig 4(c). These coefficients are more than an order of magnitude larger than those reported for the polar magnet Ni$_3$TeO$_6$ (ref. 24), see Fig 4(d).

When both external fields $H$ and $E$ are collinear and their direction coincides with the positive $c$ axis of the crystal, both $\Delta P$ and $\Delta M$ are negative in applied positive $H$ and $E$, respectively. Thus, both d$P$/d$H$ and d$M$/d$E$ are negative. The data of Figs 3(a,b) show that, consistent with exchange striction mechanism, $\Delta M$ changes sign in negative $H$, while $\Delta P$ does not. As a result, both d$P$/d$H$ and d$M$/d$E$ change their sign and become positive in negative $H$ and $E$, while retaining the same magnitudes. This sign reversal is illustrated in Supplementary Fig 4.

In conclusion, polar magnets clearly possess a great potential as ME materials. The absence of poling requirements makes possible utilization of giant ME coefficients associated with sharp metamagnetic transitions practical, because reproducible, hysteresis-free linear responses can be achieved, as necessary for applications. In Fe$_2$Mo$_3$O$_8$, hidden ferrimagnetism of the Fe-O layers strongly enhances the magnetic response at the transition field, providing explanation for the observed giant differential ME coefficients. Exchange striction mechanism of the ME effect in Fe$_2$Mo$_3$O$_8$ provides an additional functional capability of controlling the sign of these coefficients by the direction of the applied "bias" magnetic field. Therefore, studies of other polar magnets, especially with exchange striction ME mechanism and local ferrimagnetism, are, in our opinion, of significant promise.

**Methods**

**Single crystal preparation and structure analysis**

Fe$_2$Mo$_3$O$_8$ single crystals were grown using a chemical vapor transport method at 1000 °C for 10 days, followed by furnace cooling. They are black hexagonal plates with typical size ~1 × 1 × 0.5 mm$^3$. Powder X-ray diffraction measurement was performed on crushed powders of Fe$_2$Mo$_3$O$_8$ single crystals. Refinement shows that the room-temperature space group is P6$_3$mc. $a$- and $c$-lattice constants are 5.773(3) and 10.054(3) Å, respectively.

**Measurements**

All measurements of magnetic properties $M(H)$, $\chi(T)$ and $M(E)$ were performed in a Quantum Design MPMS-XL7. The dielectric constant $\varepsilon(T)$, specific heat $C_P(T)$, electric polarization $P(T)$ and $P(H)$ properties were performed using Quantum Design PPMS-9. $\varepsilon(T)$ was measured with 1V a.c. electric field applied along the $c$ axis using a Quadtech 7600 LCR meter at 44 kHZ. Specific heat measurements were conducted using the standard relaxation method. $P(T)$ and $P(H)$ were obtained by integrating the pyroelectric current $J(T)$ and magnetoelectric current $J(H)$, which were measured using Keithely 617 programmable electrometer at 5 K/min warming rate and ramping magnetic field with 200 Oe/s.

**First principles calculations**

Ab-initio calculations were performed using the full-potential linearized augmented plane wave (FP-LAPW) method as implemented in the WIEN2k code[35] within the framework of density functional theory[36,37]. The electronic, magnetic and structural properties of $Fe_2Mo_3O_8$ were calculated using the generalized gradient approximation (GGA) for the exchange-correlation potential, in the form of Perdew, Burke and Ernzerhof[38,39] (PBE) plus an on-site Coulomb self-interaction correction potential ($U$) treated by DFT+U, and the double-counting in the fully localized limit[40]. Since the symmetry of low temperature crystal structure is not known, the point group symmetry of the hexagonal paramagnetic space group P6$_3$mc (ref. 31) was artificially reduced for the purpose of optimizations of internal parameters (OIP). All the calculations were done in the triclinic space group P1, with the lattice parameters kept fixed to $a = b = 5.773$ Å, $c=10.054$ Å, $\alpha = 90°$, $\beta = 90°$, $\gamma = 120°$. OIP were performed with imposed AFM and FRM magnetic configurations, using as the initial guess the experimentally determined internal parameters of the paramagnetic phase[31]. The search for equilibrium ionic positions was carried out by means of the PORT method[41] with a force tolerance $\leq 0.5$ mRy/Bohr. The calculations were performed with more than 200 k-points in the irreducible wedge of the Brillouin zone (10 × 10 × 4 mesh). The total energy, charge and force convergence criteria were $\sim 10^{-4}$ Ry, $\sim 10^{-4}$ electrons and 0.25 mRy/Bohr, respectively. The muffin-tin radii $R_{MT}$ were chosen as 1.90, 1.93 and 1.66 bohr for Mo, Fe and O, respectively. To ensure that no charge leaks outside the atomic spheres, we have chosen the energy which separates the core and the valence states to be $-10$ Ry, thus treating the Mo(4s, 4p, 4d, 5s), Fe(3s, 3p, 3d, 4s) and O(2s, 2p) electrons as valence states. All other input parameters were used with their default values.

**Acknowledgements**

This work was supported by DOE under Grant No. DE-FG02-07ER46382. K.H. was supported by NSF-DMR 1405303. Heat capacity measurements at NJIT (T.A.T.) were supported by DOE under Grant DE-FG02-07ER46402. The PPMS (used for heat capacity measurements) was acquired under NSF MRI Grant DMR-0923032 (ARRA award). Discussions with Karin M. Rabe are gratefully acknowledged.

**Author contributions**

Y.Z.W. carried out magnetoelectric, magnetic and dielectric measurements. T.A.T. performed the heat capacity measurement. B.G. synthesized single crystals and performed XRD. G.L.P. developed the theoretical model under the supervision of K.H. Y.Z.W., G.L.P., V.K. and S.-W.C. wrote the manuscript. S.-W.C. initiated and supervised the research.

**Competing financial interests**

The authors declare no competing financial interests.

**Figure 1| Magnetic transition in $Fe_2Mo_3O_8$.** (a) Crystal structure of $Fe_2Mo_3O_8$. Vertical lines connect the nearest Fe ions along the *c* axis (blue lines are longer than the red ones). (b) The Fe-O layer in the *ab* crystallographic plane. Thick line depicts the largest Fe-Fe magnetic coupling *J*. (c) Schematic view of the AFM and FRM orders. Pink arrows represent the ferrimagnetic moments of the individual Fe-O layers. (d) The AFM order, together with the calculated largest ionic shifts associated with the paramagnetic to AFM transition. The direction of the magnetically-induced $\Delta P$ is shown with a thick arrow. (e) Temperature dependence of DC magnetic susceptibility $\chi_{DC}$ in zero field-cooled (ZFC) and field-cooled (FC) processes along two crystallographic directions, parallel and perpendicular to the *c* axis, in $\mu_0 H$=0.2 T. (f) Specific heat anomaly at the Neel temperature. Red line represents the double Debye model fit discussed in the text. Insert: the image of as-grown $Fe_2Mo_3O_8$ single crystal.

**Figure 2| Magnetically-induced electric polarization, and the metemagnetic transition.** (a) Temperature dependence of the *c*-axis dielectric constant $\epsilon(T)$, *f*=44 kHZ. (b) Variation of the *c*-axis electric polarization $\Delta P$ with temperature. (c,d) Magnetic field dependence of magnetization *M(H)* and polarization $\Delta P(H)$ at various temperatures. In (d), solid (open) circles depict the data obtained upon sweeping the magnetic field up (down).

**Figure 3| Magnetoelectric effect, and the associated ionic shifts.** (a,b) Magnetic field dependence of magnetization *M(H)* and polarization *P(H)* at *T*=55 K. Numbers and arrows indicate the measurement sequence. The insert in (b) shows the magnetic orders and the ferrimagnetic moments of the Fe-O layers for the phases involved. (c) The calculated ionic shifts for the paramagnetic to AFM transition. Thick arrow represents the corresponding change of the electric polarization, $\Delta P$. (d) the same as (c), but for the AFM to FRM transition.

**Figure 4| Reproducible magnetoelectric control of the electric polarization and magnetization with giant ME coefficients.** (a) Periodic modulation of electric polarization (blue) induced by a magnetic field linearly varying between 3.25 T and 3.5 T (black) at 55 K. (b) Periodic modulation of magnetization (green) induced by an electric field (red) linearly varying between ±16.6 kV/cm, for *T*=55 K and $\mu_0 H$ =3.345 T. (c) Phase diagram of $Fe_2Mo_3O_8$. Black dots determined from *M(H)*, and red diamonds – from $\chi(T)$ curves. (d) Electric field dependence of magnetization for $Fe_2Mo_3O_8$ (from panel (b), averaged), and for $Ni_3TeO_6$ (×10). The insert illustrates the experimental setup, with directions of the applied fields shown. In all figures, the magnetization, polarization, and the applied fields are along the *c* axis.

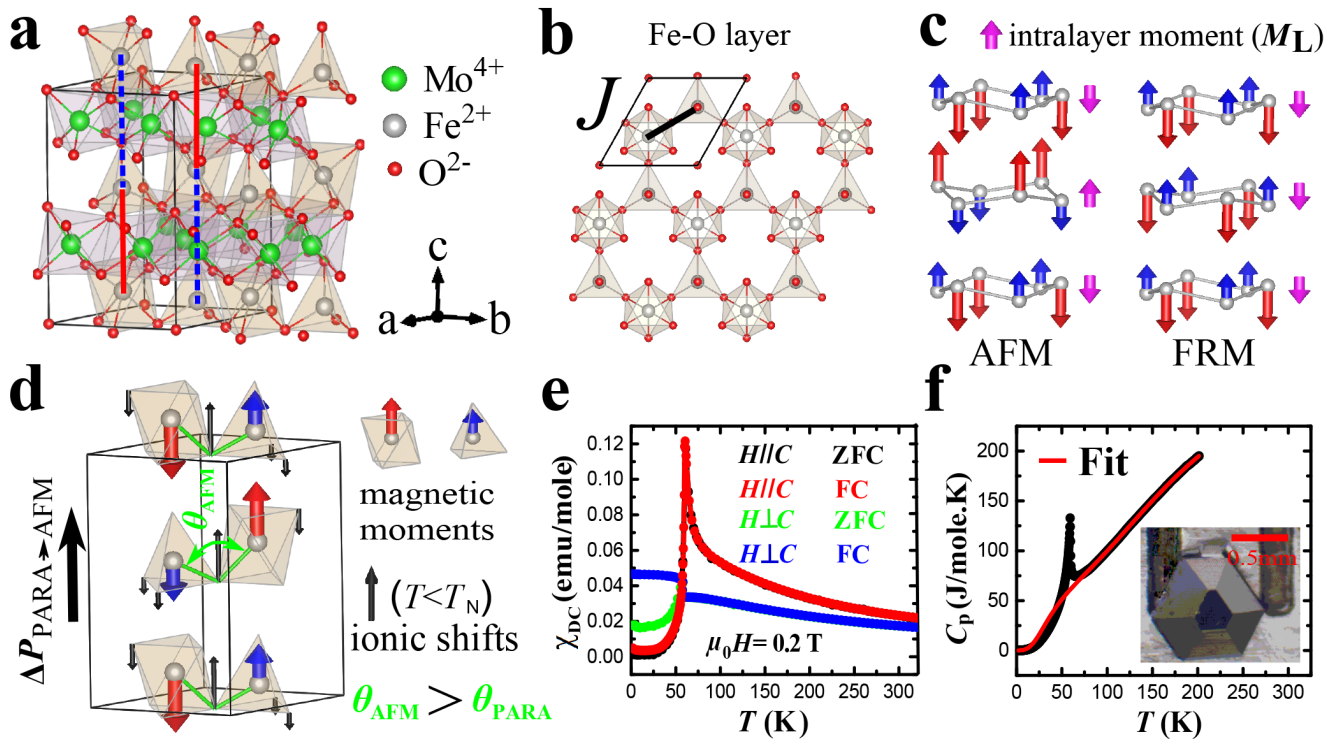

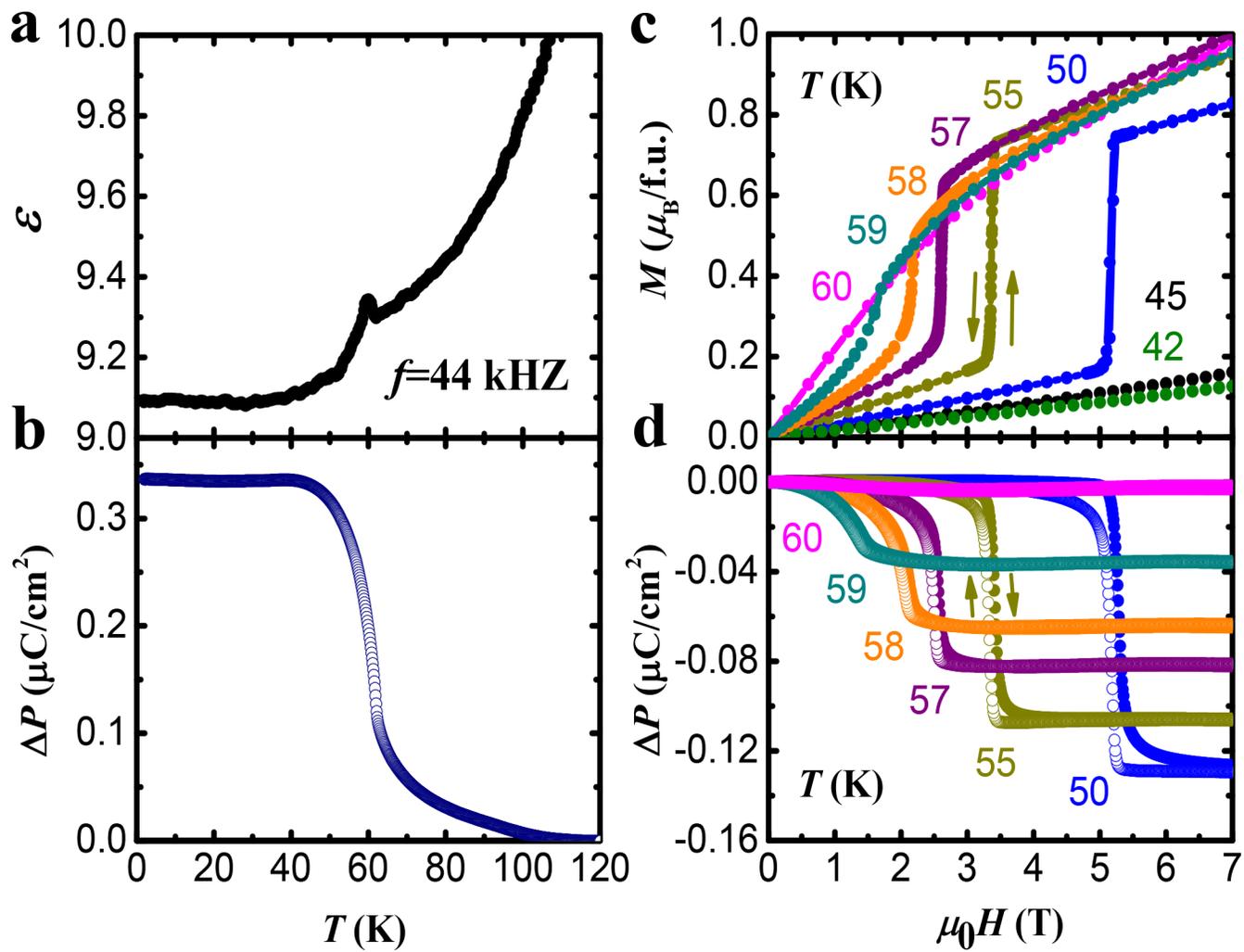

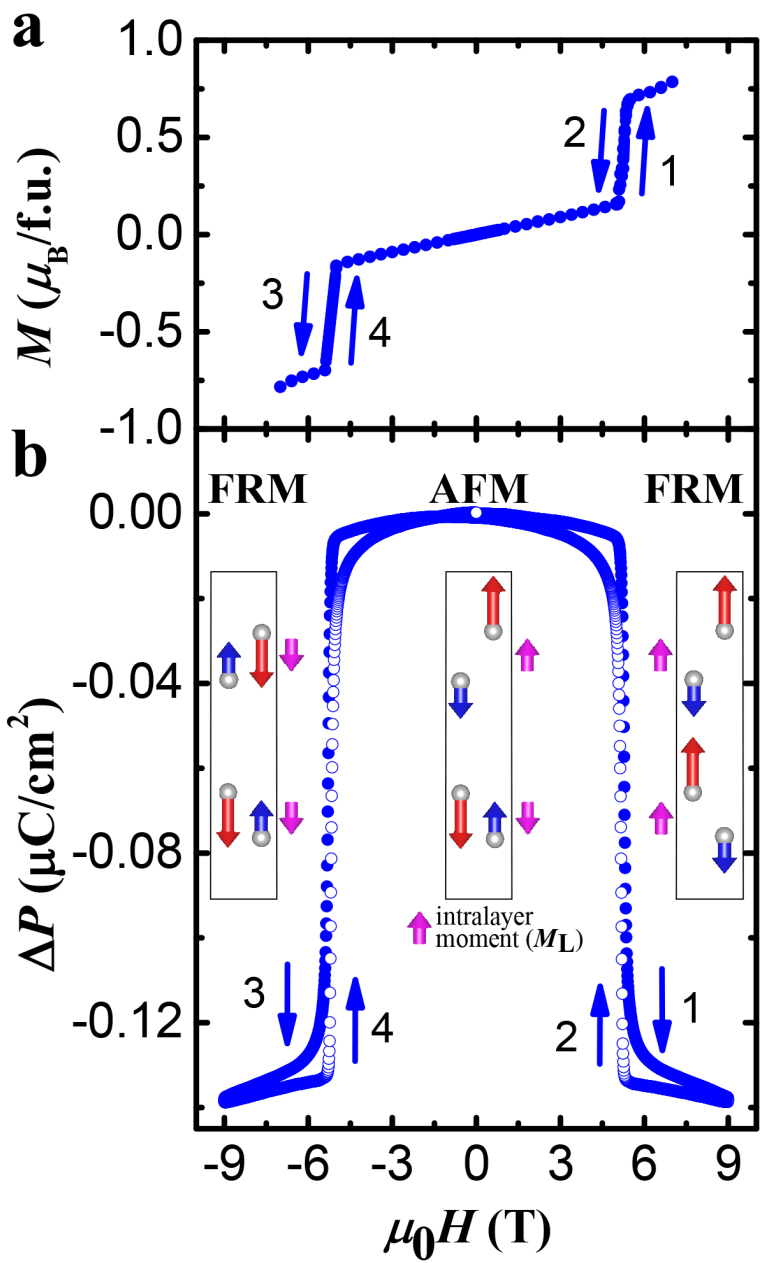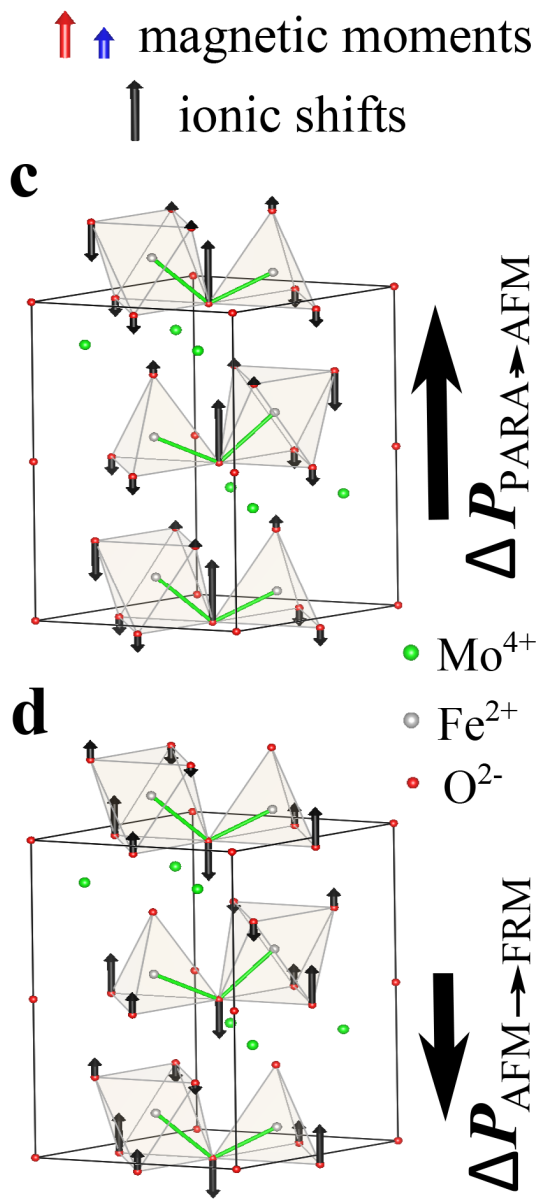

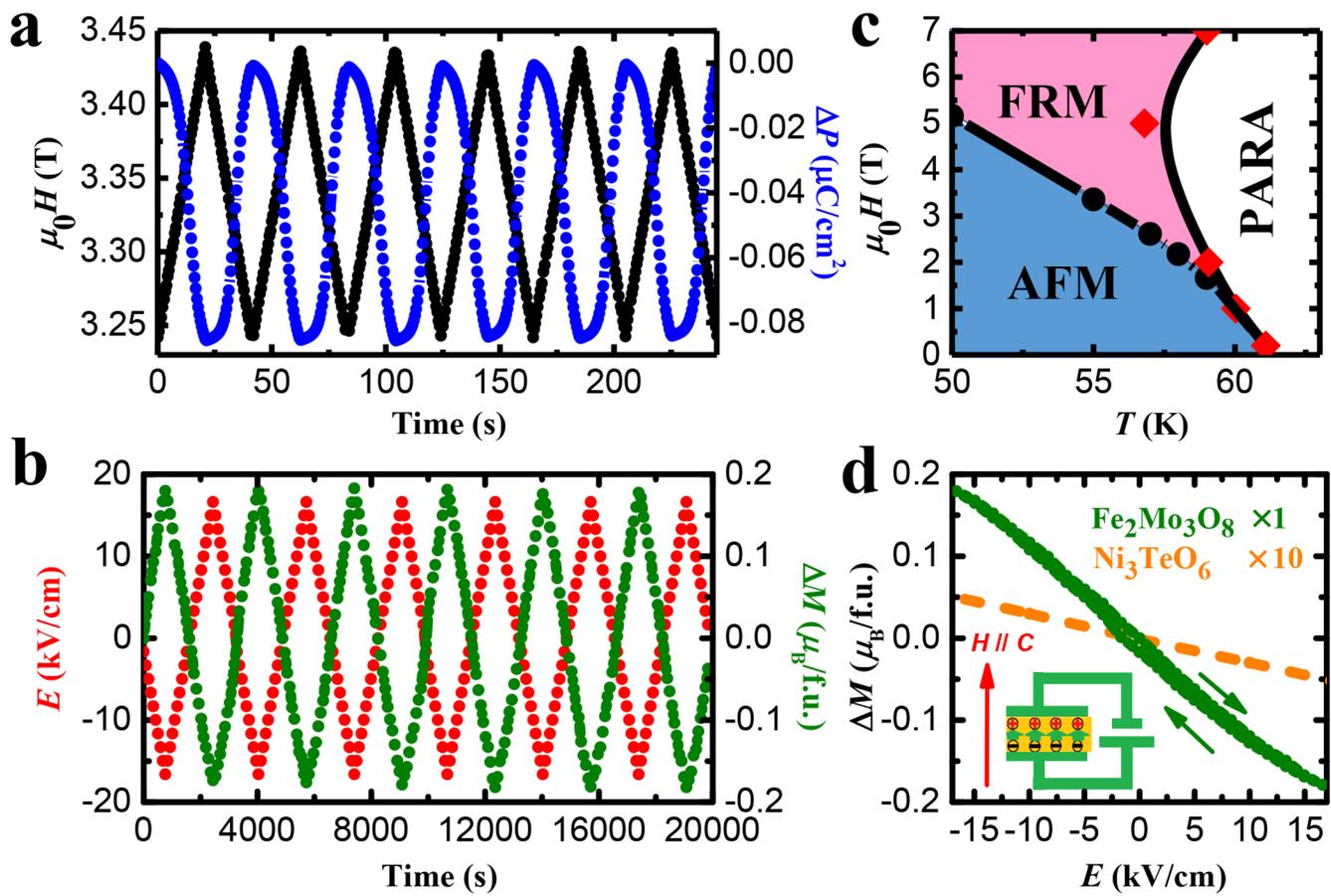